\documentstyle[prb,aps]{revtex}
\newcommand{\be}{\begin{equation}} \newcommand{\ee}{\end{equation}}
\newcommand{\ba}{\begin{eqnarray}} \newcommand{\ea}{\end{eqnarray}}
\newcommand{\bit}{\begin{itemize}} \newcommand{\eit}{\end{itemize}}
\newcommand{\ben}{\begin{enumerate}} \newcommand{\een}{\end{enumerate}}

 \newcommand{\dpa}{\partial}

\begin{document}
\large

\title{\Large Landau damping: is it real? }
\author{V.~N.~Soshnikov                                          %
 \thanks{Correspondence should be addressed to: Soshnikov~V.~N., %
         Krasnodarskaya str., 51-2-168, Moscow 109559, Russia.}  %
        }
\address{%
Plasma Physics Dept.,
All-Russian Institute of Scientific and Technical Information\\
of the Russian Academy of Sciences
(VINITI, Usievitcha 20, 125219 Moscow, Russia)}%
\maketitle
  \vspace{-0.13\textheight}

   \hspace*{0.44\textwidth} 
    \underline{\bf http://xxx.lanl.gov/e-print/physics/9712013}\\
     \vspace{0.09\textheight}

\begin{abstract}
  To calculate linear oscillations and waves in dynamics of gas
and plasma one uses as a rule the old classical method 
of dispersion equation. 
The latter connects frequencies $\omega$ and wave numbers $k$
(both are in general case complex values): 
 $\epsilon(\omega,k)=0$. In the plasma case $\epsilon$ is called
generalized dielectric susceptibility.
Dispersion equation is derived by substitution of 
asymptotical solution $\exp\left(-i\omega t+i\vec k\vec r\right)$
into the coupled system of linearized dynamical and field 
partial differential equations.
   However, this method appears to be inapplicable, f.e.,
in the case of waves in Maxwellian collisionless plasma
when dispersion equation has no solutions.
By means of some refined sophistication L.~Landau in 
1946~\cite{bib-1} has suggested in this case actually 
to replace the dispersion equation with another one, 
having a specific solution (``Landau damping'')
and being now widely used in plasma physics.
   Recently we have suggested~\cite{bib-13} a quite new
universal method of two-dimensional Laplace transformation 
(in coordinate $x$ and time $t$ for plane wave case),
that allows to obtain asymptotical solutions 
of original Vlasov plasma equations as inseparable sets
of coupled oscillatory modes 
(but not a single wave like $\exp\left(-i\omega t+i k x\right)$).
The mode parameters are defined in this case by double-poles
($\omega_n$,$k_n$) of Laplace image $E(\omega_n,k_n)$ 
of electrical field $E(x,t)$ or, correspondingly,
by double-zeros of inverse expression $1/E(\omega_n,k_n)=0$.
This method, in contrast with classical one,
allows to obtain the whole set of oscillatory modes
for every concrete problem.
It leads to some new ideology in the theory of 
plasma oscillations, 
the latter's being a set of coupled oscillatory modes
(characterized by pairs ($\omega_n$,$k_n$) and amplitudes) 
and depending not only on the intrinsic plasma parameters,
(as is the case in classic theory), 
but also on mutually dependent self-consistent initial 
and boundary conditions 
and on method of plasma oscillations excitation.
\end{abstract}
\vspace{0.7mm}

\vspace{0.7mm}


\hrule
\vspace*{5mm}

  More than 50 years ago there appeared the work of
L.~Landau~\cite{bib-1} with an original asymptotical solution 
of Vlasov equations for collisionless Maxwellian plasma 
in the form of a single travelling damping electromagnetic wave
(longitudinal or transversal one). 
In this way there appeared some novel conceptions in plasma physics:
Landau damping and Landau rule of bypassing around poles
in calculating some indefinite Cauchy-type integrals 
along the real axis, which arise in standard procedures of
deriving dispersion equation after substituting in Vlasov equations
the wave $\exp\left(-i\omega t + ikx\right)$. 
Landau searched for a solution $f(t)$ of initial problem 
for perturbation of electron distribution function and
electrical field in a form $\sim\exp(ikx)\cdot f(t)$,
where $k$ is real wave number, 
using the method of Laplace transformation of $f(t)$.
By means of analytical continuation of Laplace image of $f(t)$,
that is $f_p$, in complex transform parameter $p\equiv -i\omega$
he has obtained an asymptotical solution corresponding to 
a pole $p_n$ of the analytically continued function 
in the form of a slowly damping travelling wave 
with complex frequency $\omega\equiv\omega_0-i\delta$.
This work of Landau was attributed to the rank of outstanding
discoveries in the physics of plasma~\cite{bib-2}.

   The most surprising turned out the fact that 
the substitution of this solution back into the Vlasov equations
results in the dispersion equation which has no solutions 
at $\omega_0\neq 0$ for any $\delta$ 
(see in more details~\cite{bib-3,bib-4}).
Yet, it was decided that the right one is not 
the dispersion equation obtained in this way,
but some {\em other} dispersion equation 
with evident natural additions in the Vlasov equations.
These additions arise at calculation of indefinitely divergent
Cauchy integral in velocity $v_x$ with the pole $\omega=kv_x$ 
($x$ is direction of wave movement, $v_x$ is $x$-component
of particle velocity and integration along real $v_x$-axis
is carried out with bypassing this pole along the half-circle
contour in the complex plane $v_x$).

  These results came into the all plasma physics text-books
(see, f.e.~\cite{bib-5,bib-6,bib-7,bib-8}). There have appeared
experimental works~\cite{bib-9,bib-10,bib-16,bib-17,bib-18},
which seemed to have proved Landau damping.
Although sometimes there appeared some doubts about rightness
of these theoretical results~\cite{bib-11,bib-12,bib-14,bib-15},
all of them were searching for either
something mathematically incorrect in Landau's derivation
or some diversifications, and also philosophical justifications
like the causality principle or expansion in asymptotically
divergent series in small ratio $\delta/\omega_0$, and so on.
To date Landau's results are considered as irrefragable,
and they compose an essential part of the theoretical plasma
physics~\cite{bib-2,bib-5,bib-6,bib-7,bib-8}.
Attempts to find mathematical incorrectnesses in Landau's
derivation, in particularly, in~\cite{bib-3,bib-4},
have failed (see~\cite{bib-13}). Apparently, 
his derivation is really mathematically irreproachable.

  These results are looked now in a quite another light after
the successful attempt to use in solving Vlasov equations 
two-dimensional (in coordinate $x$ and time $t$)
Laplace transformation~\cite{bib-13,bib-3,bib-4}.
In this case the asymptotical solution {\em which must satisfy}
the original Vlasov equations after its substitution there,
is found as {\em the sum of oscillatory modes} 
$\sum_{n}a_n\exp\left(-i\omega_nt+ ik_nx\right)$
which does not reduce to the simple product $f(t)\phi(x)$.
And besides, coefficients $a_n$, frequencies and wave numbers
being found not from the dispersion equation 
$\epsilon(\omega,k)=0$~\cite{bib-6,bib-7}
($\epsilon$ is generalized plasma susceptibility),
but as some values determining residues and double-poles 
of the Laplace transform $E(\omega,k)$ or, correspondingly,
double zeros (in $\omega$ and $k$) of the inverse expression
$1/E(\omega,k)$.
In this case the solution, including $\omega_n$, $k_n$,
is determined not only by inherent parameters of plasma
(as it is in the Landau theory), 
but in the same extent by initial as well as boundary conditions
and by the ways of excitation of plasma oscillations/waves.
The initial and boundary conditions are not arbitrary,
but are connected by conditions of the field self-consistency
(supposing finiteness) in Vlasov equations, 
so a dividing to the so called properly initial 
or properly boundary problems in Landau theory is incorrect.
In this way, even at the initial condition of perturbation 
of the distribution function 
$f_1\left(x,t=0,\vec v\right)=
 \alpha\left(\vec v\right)\exp(ikx)$
initial single-mode solution is factorized in time, 
and other oscillatory modes appear inevitably
in the asymptotical solution.
The coupled modes which are constituents of the solution
include Langmuir waves, standing waves, 
damping and non-damping waves and oscillations, but
{\em do not include whatever specific Landau damping}~\cite{bib-13}.
Vlasov equations must be satisfied indeed only 
by the totality of the modes of $f_1\left(x,t,\vec v\right)$
and $\vec E(x,t)$,
but not by whatever mode separately.
Besides that, from the condition of self-consistency (finiteness)
of solution it follows that the modes of $f_1$ are proportional
to the amplitude of electrical field with the natural result 
that there are no oscillations in the absence of restoring forces
of electrical field (when the latter is damping).
Properties of the system of linear equations
which connect asymptotical amplitudes of $E(x,t)$ and 
$f_1\left(x,t,\vec v\right)$ determine the form of solution:
one, or more independent sets of modes including 
a possible case of unavoidable exponentially divergent modes 
(probably, it is the peculiar case of an external
non-self-consistent source).

  Strictly speaking, in order to keep self-consistency
in advance, one could after setting boundary and initial
conditions for the distribution function $f_1$ 
also obtain such conditions for electric field $E$
in accord with Poisson equation for $\dpa E/\dpa x$.
\vspace{2mm}

 {\Large \bf  Whether these results being contrary to all 
the educational text-books on plasma physics are right?}
\vspace{2mm}

  The answer is, that Landau derivation appears only 
as some mathematically right part of the classical proof 
by {\em reductio ad absurdum}:
from some arbitrary supposition, what would be,
if the solution of Vlasov equations should have the form
$f(t)\exp(ikx)$, follows at asymptotical limit 
$f(t)\to\exp(-\omega t)$ with Landau's  $\omega$, $k$ 
(and also at any other values), 
that these solutions do not satisfy the original Vlasov equations,
{\em consequently}, the initial supposition about 
the assumed form of the solution was wrong. 
It is a strikingly simple and classical solution 
of the ``paradox of Landau'', 
including some mysterious appearance of non-damping oscillations of
$f_1\left(x,t,\vec v\right)$ in the absence of any restoring force
(electrical field $E\to 0$ according to Landau damping!)~\cite{bib-5}.

   However the contradiction in Landau's logic is even more profound.
He proceeds from the Laplace image $f_p^+$ 
(correspondingly, electrical field $E_p^+$) in the upper half-plane
$\omega$, where poles in $\omega$ should lead 
to exponentially divergent solutions, what appears illogical. 
But in the upper half-plane poles are absent, 
so the dispersion equation has no solutions.
But Landau's analytical continuation in $\omega$ 
into the lower half-plane leads to poles in $\omega$,
which contrary to the initial supposition correspond 
to exponentially damping solutions (what already is paradoxical).
Besides, these solutions satisfy some another dispersion equation
(the original dispersion equation which follows from 
substitution of a travelling wave into Vlasov equations,
as it was already said before, is not satisfied in this case).
On the contrary, it should be more logical
to proceed from the Laplace images in the lower half-plane
$f_p^-$, $E_p^-$ (different from $f_p^+$, $E_p^+$)
with damping solutions.
However, in this case the poles exist only for 
analytical continuation of $f_p^-$, $E_p^-$ into the upper half-plane
and correspond, contrary to the initial supposition,
to divergent solutions.

   So, our basic principle is to find the correct solution 
of original Vlasov equations, but not to find some modified equations
for fitting them to be satisfied by Landau damping solution.

   In this way, the dispersion equation in the form
      $\epsilon(\omega,k)=0$,
as well as generally the method of one-mode derivation
of dispersion equation have rather limited applicability.
The suggested here method of two-dimensional Laplace transformation
gives effectively the total set of coupled oscillatory modes 
in every concrete case, 
and it can lead to many new results not only in plasma physics
(with revision of the results obtained on the base of Landau
``bypassing around poles''), but also in gasdynamics and other fields.
One can also conclude that with the new method 
a completely new ideology of oscillatory solutions arises 
which is quite different with respect to Landau's 
sophisticated conceptions, ``predicting'' physical phenomenon
(``Landau damping''), which probably does not exist in nature.

  Experimental ``Landau damping'' 
(see~\cite{bib-2,bib-9,bib-10,bib-16,bib-17,bib-18}) may be mimicked
by a damping caused by finiteness of plasma tube diameter
(in the real plasma slab modelling) or/and non-selfconsistent
(external) field term of the plasma waves excitation
(in the field term of kinetic equation),
or more really by a non-zero source term in the right-hand side
of the more general kinetic equation(s) (cf.~\cite{bib-2}),
according to some concrete experimental conditions
(Langmuir probes type, excitation transparent grids 
or compact flat or wire electrodes etc.). 
So, this experimental damping must satisfy some other physical
equations, not the original Vlasov equations.
In this case the actual problem is the analysis 
and physical groundings of other different modifications
of these equations in their conformity with the different concrete
experimental conditions, 
including a distinct demarcation between the electric external
(non-self-consistent) and intrinsic Poissonian
(self-consistent) fields.
Such plasma wave equations can be solved then with 
the proposed method of two-dimensional Laplace transformation.

\acknowledgements

  I am very grateful to Dr.~A.~P.~Bakulev for his active
and very constructive edition and help in amendment of this paper
as well as his realization of electronic sending procedures.

\end{document}